\documentclass[fleqn,usenatbib,useAMS]{mnras}

\usepackage{newtxtext,newtxmath}
\usepackage[T1]{fontenc}
\usepackage{ae,aecompl}
\usepackage{graphicx}	
\usepackage{amsmath}	
\usepackage{xspace}

\usepackage{xcolor}
\usepackage{hyperref}

\newlength\q
\usepackage{array}
\newcolumntype{C}[1]{>{\centering\let\newline\\\arraybackslash\hspace{0pt}}m{#1}}

\newcommand{\be}{\begin{equation}}
\newcommand{\ee}{\end{equation}}
\newcommand{\bea}{\begin{eqnarray}}
\newcommand{\eea}{\end{eqnarray}}
\newcommand{\no}{\nonumber}

\newcommand{\Mpeakz}{\ensuremath{M_{\rm peak}^{z=0}}\xspace}

\newcommand{\Mpeakzall}{\ensuremath{M_{\rm peak}^{z}}\xspace}

\definecolor{mypink2}{RGB}{219, 48, 122}

\title[The dependence of assembly bias on the cosmic web]
{The dependence of assembly bias on the cosmic web}

\author[Montero-Dorta et al.]{
\parbox[t]{\textwidth}{
Antonio D. Montero-Dorta$^{1}$\thanks{E-mail: amonterodorta@gmail.com} \& Facundo Rodriguez$^{2,3}$} 
\vspace*{6pt} \\ 
$^{1}$Departamento de F\'isica, Universidad T\'ecnica Federico Santa Mar\'ia, Casilla 110-V, Avda. Espa\~na 1680, Valpara\'iso, Chile.\\
$^{2}$ CONICET. Instituto de Astronom\'ia Te\'orica y Experimental (IATE). Laprida 854, X5000BGR, C\'ordoba, Argentina. \\
$^{3}$ Universidad Nacional de C\'ordoba (UNC). Observatorio Astron\'omico de C\'ordoba (OAC). Laprida 854, Córdoba X5000BGR, Argentina.\\
\vspace{-0.4cm} 
}

\date{Accepted ---. Received ---;in original form --- \vspace{-0.3cm}}

\pubyear{2021}

\def\simlt{\lower.5ex\hbox{$\; \buildrel < \over \sim \;$}}
\def\simgt{\lower.5ex\hbox{$\; \buildrel > \over \sim \;$}}

\definecolor{red}{rgb}{1,0,0}

\begin{document}
\label{firstpage}
\pagerange{\pageref{firstpage}--\pageref{lastpage}}
\maketitle


\begin{abstract}

For low-mass haloes, the physical origins of halo assembly bias have been linked to the slowdown of accretion due to tidal forces, which are expected to be more dominant in some cosmic-web environments as compared to others. In this work, we use publicly available data from the application of the Discrete Persistent Structures Extractor (DisPerSE) to the IllustrisTNG magnetohydrodynamical simulation to investigate the dependence of the related galaxy assembly bias effect on the cosmic web. We first show that, at fixed halo mass, the galaxy population displays significant low-mass secondary bias when split by distance to DisPerSE critical points representing nodes ($d_{\rm node}$), filaments ($d_{\rm skel}$), and saddles ($d_{\rm sadd}$), with objects closer to these features being more tightly clustered. The secondary bias produced by some of these parameters exceeds the assembly bias signal considerably at some mass ranges, particularly for $d_{\rm sadd}$. We also demonstrate that the assembly bias signal is reduced significantly when clustering is conditioned to galaxies
being close or far from these critical points. The maximum attenuation is measured for galaxies close to saddle points, where less than 35$\%$ of the signal remains. Conversely, objects near voids preserve a fairly pristine effect (almost 85$\%$ of the signal). Our analysis confirms the important role played by the tidal field in shaping assembly bias, but they are also consistent with the signal being the result of different physical mechanisms. Our work introduces some new aspects of secondary bias where the predictions from hydrodynamical simulations can be directly tested with observational data.

\end{abstract}


\section{Introduction}
\label{sec:intro}

It is well established that the large-scale linear bias of dark-matter (DM) haloes depends strongly on their internal properties. Among these properties, halo mass is responsible for the primary dependence, as a direct manifestation of the intrinsic dependence of bias on the peak height of density fluctuations, $\nu$. More massive haloes are more tightly clustered than less massive haloes, in agreement with the $\Lambda$-cold dark matter ($\Lambda$-CDM) structure formation formalism (e.g., \citealt{Press1974,ShethTormen1999,Sheth2001,ShethTormen2002}). In recent years, a number of additional {\it{secondary dependencies}} at fixed halo mass have been unveiled using cosmological simulations (see, e.g., \citealt{Sheth2004,gao2005,Wechsler2006,Gao2007,Dalal2008, Angulo2008,Li2008,faltenbacher2010, Lazeyras2017,2018Salcedo,han2018,Mao2018, SatoPolito2019, Johnson2019, Ramakrishnan2019,MonteroDorta2020B, Tucci2021, MonteroDorta2021}). The one that has drawn more attention is the dependence on the assembly history of haloes, an effect commonly referred to as {\it{halo assembly bias}}. Lower mass haloes that assemble a significant portion of their mass early on are more tightly clustered than haloes that form at later times, with the signal progressively vanishing towards the high-mass end\footnote{In the context of assembly bias / secondary bias, the linear bias is usually computed on scales of 5-20 $h^{-1}$Mpc.} (e.g., \citealt{gao2005, Li2008, SatoPolito2019, MonteroDorta2021}). 

It is reasonable to expect that these secondary halo clustering dependencies have a direct impact on the central galaxy population; at least, as much as the scatter in the stellar-to-halo mass relation (SHMR) permits (\citealt{Guo2010,Behroozi2013,Wechsler2018}). In this context, a simple way to define {\it{galaxy assembly bias}} is the dependence of galaxy clustering on the assembly history of haloes at fixed halo mass. Other different but related definitions are, however, common in the literature, including those based on halo {\it{occupancy variations}} (e.g., \citealt{Artale2018,Zehavi2018,Bose2019,Salcedo2022}) or so-called {\it{anisotropic assembly bias}} (e.g., \citealt{Obuljen2019,Obuljen2020}). Although these effects have been measured in hydrodynamical simulations (\citealt{Artale2018,MonteroDorta2020B, MonteroDorta2021}), providing a broadly accepted proof of their existence using observational data has been quite challenging. Despite some claims, no consensus has been reached in terms of the actual detectability of either galaxy assembly bias or the underlying secondary halo bias effects (see, e.g., \citealt{Miyatake2016, Zu2016,Lin2016,MonteroDorta2017B,Niemiec2018,Obuljen2019,Obuljen2020,Salcedo2022,Wang2022,Sunayama2022}). This uncertainty, along with the potential repercussion of assembly bias on cosmological measurements, has turned the attention of the community towards this field in recent years.

From the galaxy evolution perspective, the fact that galaxies (and DM haloes) are influenced by the properties of the environment where they reside has been known and studied for decades \citep[e.g.,][]{Dressler1980, Heavens1988,Kauffmann2004,Avila2005, Libeskind2013,Das2015,Poudel2017,Ganeshaiah2019, Hellwing2021, Alfaro2020, Alfaro2021}. Hierarchical growth of structure dictates that the matter distribution in the Universe forms a {\it{cosmic web}}, i.e., a complex network of {\it{filaments}} and {\it{sheets}} connecting {\it{nodes}}, and separated by empty regions or {\it{voids}}. In the visible Universe, the high-density nodes of the web correspond to galaxy groups and clusters, connected by elongated filaments and walls of galaxies and flowing gas, which in turn surround regions where galaxies are scarce (e.g., \citealt{Bond1996, ForeroRomero2009, Aragon2010,Cautun2014,Tempel2014, Weygaert2016}). In recent years, the characterisation of the cosmic well has become progressively more precise, which has allowed more detailed analyses of the dependence of the properties of gas and galaxies on their specific locations within these structures \citep[e.g.,][]{Porter2008,Codis2012,Tempel2013, Cautun2014, Laigle2015, Kraljic2018, Ganeshaiah2019, Pereyra2020, Rost2021}.

One of the most innovative techniques to identify the structures making up the cosmic web is based on Morse's theory \citep{Morse1934}. This approach characterises the spatial connectivity of the smoothed density field in terms of a set of critical points; mainly maxima, minima and inflexion points. This procedure has lead to the development of powerful tools to outline the ``skeleton" of the Universe. Among these tools, the Discrete Persistent Structures Extractor \citep[DisPerSE; ][]{Sousbie2011a,Sousbie2011b} provides a sophisticated formalism that facilitates the detection of cosmic-web structures by adding so-called persistence analysis to topological identification. In this paper, we use publicly available DisPerSE data to analyse the dependence of galaxy assembly bias on the location of galaxies (and haloes) within the cosmic web (or more precisely, on the distance to the critical points provided by the DisPerSE description). We employ the DisPerSE data provided as a value-added catalogue within the IllustrisTNG\footnote{\url{http://www.tng-project.org}} magnetohydrodynamical simulation. Projects such as TNG have become essential tools to gain insight into a variety of aspects concerning the halo-galaxy connection. Regarding secondary halo bias, \cite{MonteroDorta2020B, MonteroDorta2021} used TNG to demonstrate that secondary halo bias manifests itself on the central galaxy population when this is split by several galaxy properties at fixed halo mass. 

The link between assembly bias (or secondary bias, more generally) and the properties of the local environment has been addressed previously from several perspectives. \cite{Dalal2008} introduced the notion that low-mass halo assembly bias, as opposed to its high-mass counterpart, may be due to a subpopulation of haloes whose accretion was truncated early on. Subsequently, \cite{Hahn2009} analysed this suppressed growth in more detail, claiming that tidal effects produced by a neighbouring massive halo could be the dominant driver. The dependence of the formation epoch on environment density is viewed in these works as a secondary effect induced by the enhanced density of haloes in filaments near massive haloes where the tides are strong. The importance of the tidal field is also highlighted in the more recent works of \cite{Borzyszkowski2017} and \cite{Musso2018}. The results of \cite{Borzyszkowski2017}, based on zoom-in simulations of a handful of haloes, are consistent with the idea that low-mass assembly bias is associated with the existence of same-mass subpopulations of ``stalled" and ``accreting" haloes, typically living, respectively, in filaments and nodes. \cite{Musso2018}, on the other hand, derived a model based on an excursion set approach that predicts that, at fixed halo mass, mass accretion rate and formation time vary with orientation and distance from saddle points in the density field (critical points inside filaments that sit between two density peaks). Using saddles as reference points, the model shows that the large-scale bias of haloes at fixed halo mass is influenced by the geometry of the tides. 

A convenient way of looking into the connection between assembly bias and environment in a simulation catalogue is by measuring the tidal tensor around haloes; in particular, the anisotropy parameter, $\alpha$, that can be derived from it (\citealt{Paranjape2018}; the higher $\alpha$, the more anisotropic the local environment). In a series of papers, some of them using IllustrisTNG data, it has been shown that the secondary bias signals correlate with $\alpha$ at fixed halo mass (see, e.g., \citealt{Paranjape2018, Ramakrishnan2019}). Note that this parameter has also been discussed in the context of subhalo abundance matching (SHAM) techniques (e.g., \citealt{Favole2021}). 

From a variety of works, therefore, it has become clear that the secondary halo bias measured in simulations is closely tied to the tidal field, although the details of this connection and the particular physical mechanisms involved are still not fully characterised\footnote{See also \cite{Mansfield2020,Tucci2021} for discussion on related mechanisms such as the effect of splashback haloes on secondary bias.}. The DisPerSE description employed in this work offers the opportunity to add a different but complementary angle to the study of the dependence of assembly bias on the cosmic web. The paper is organised as follows. Section \ref{sec:sims} provides a brief description of the simulation data used in this work. The characterisation of the cosmic-web based on the distance to the critical points of the density field provided by DisPerSE is explained in Section \ref{sec:environment}. The main results of our analysis, in terms of the dependence of assembly bias on the cosmic web, are presented in Section \ref{sec:ABCW}. Finally, Section \ref{sec:discussion} is devoted to discussing the implications of our results and providing a brief summary of the paper. The IllustrisTNG300 simulation adopts the standard $\Lambda$CDM cosmology \citep{Planck2016}, with parameters $\Omega_{\rm m} = 0.3089$,  $\Omega_{\rm b} = 0.0486$, $\Omega_\Lambda = 0.6911$, $H_0 = 100\,h\, {\rm km\, s^{-1}Mpc^{-1}}$ with $h=0.6774$, $\sigma_8 = 0.8159$, and $n_s = 0.9667$.


\begin{figure*}
	\includegraphics[width=2.0\columnwidth]{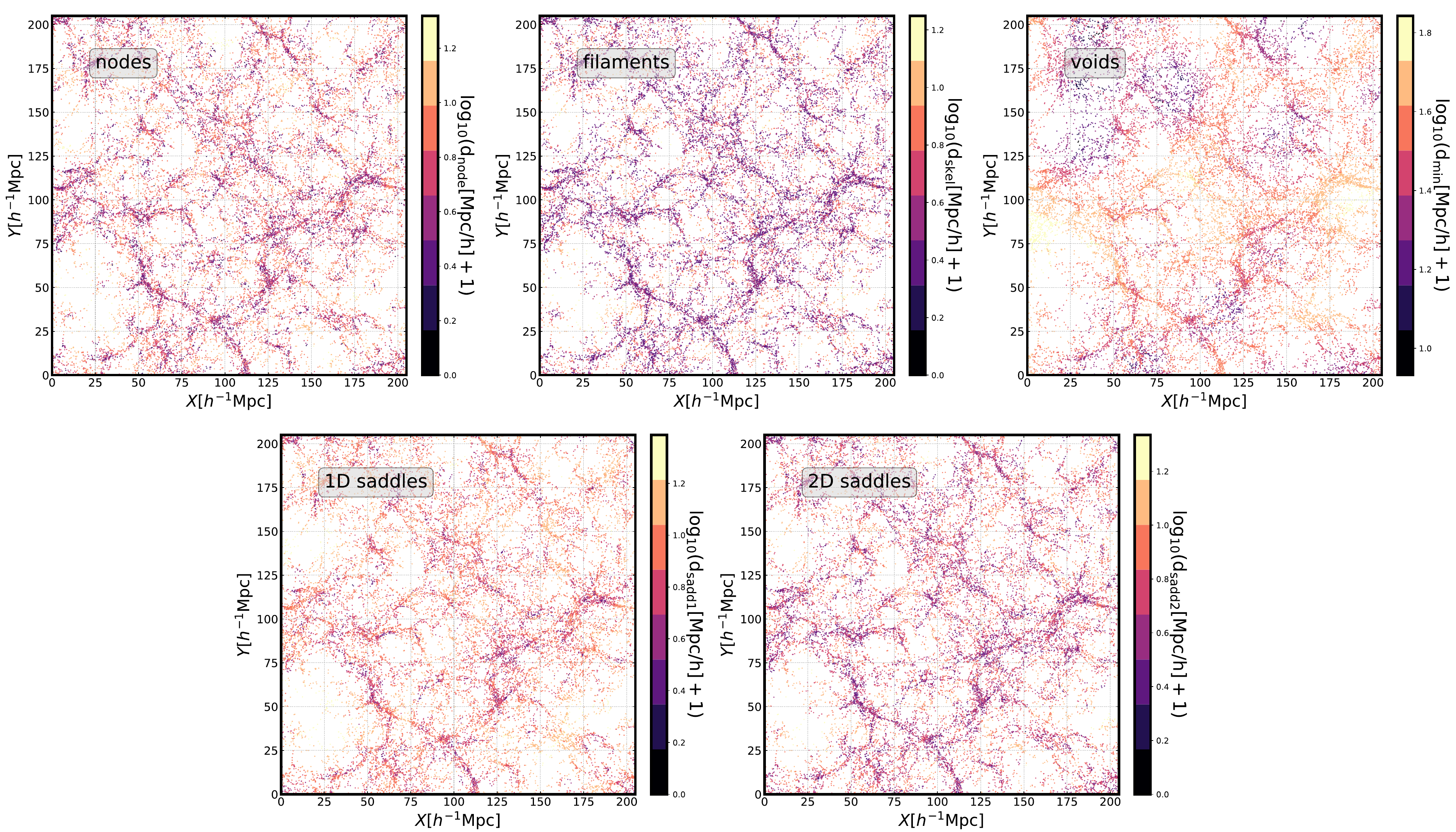}
    \caption{The cosmic web in a slice of 30x205X205 $h^{-3}$Mpc$^3$ as described by the DisPerSE parameters. In the upper row, from left to right, the maps are colour-coded according to distance to nodes ($d_{\rm node}$), filaments ($d_{\rm skel}$), and voids ($d_{\rm min}$), respectively. In the lower row, the colour code indicates distance to 1D saddle points ($d_{\rm sadd1}$) and 2D saddle points ($d_{\rm sadd2}$), respectively.}
    \label{fig:maps}
\end{figure*}

\section{Simulation data: the TNG300 box}
\label{sec:sims}

Our analysis is based on data from the IllustrisTNG magnetohydrodynamical cosmological simulation (hereafter TNG for simplicity, \citealt{Pillepich2018b,Pillepich2018,Nelson2018_ColorBim,Nelson2019,Marinacci2018,Naiman2018,Springel2018}). The TNG simulation suite was produced using the {\sc arepo} moving-mesh code \citep{Springel2010} and is considered an improved version of the previous Illustris simulation \citep{Vogelsberger2014a, Vogelsberger2014b, Genel2014}. The updated TNG sub-grid models account for star formation, radiative metal cooling, chemical enrichment from SNII, SNIa, and AGB stars, and stellar and super-massive black hole feedback. These models were calibrated to successfully reproduce a set of observational constraints that include the observed $z=0$ galaxy stellar mass function, the cosmic SFR density, the halo gas fraction, the galaxy stellar size distributions, and the black hole -- galaxy mass relation (we refer the reader to the aforementioned papers for more information).

We analyse the largest box available in the database, TNG300-1 (hereafter TNG300\footnote{\url{https://www.tng-project.org/data/docs/specifications/}}), which provides an obvious advantage when it comes to measuring large-scale halo/galaxy clustering. TNG300 spans a side length of $205\,\,h^{-1}$Mpc and includes periodic boundary conditions. The TNG300 run followed the dynamical evolution of 2500$^3$ DM particles of mass $4.0 \times 10^7$ $h^{-1} {\rm M_{\odot}}$ and (initially) 2500$^3$ gas cells of mass $7.6 \times 10^6$ $h^{-1} {\rm M_{\odot}}$. This box is a proven tool for studies of galaxy formation and the halo--galaxy connection that has provided important insight for a variety of science cases 
(see a few examples in \citealt{Springel2018,Pillepich2018,Bose2019,Beltz-Mohrmann2020,Contreras2020,Gu2020,Hadzhiyska2020,Hadzhiyska2021,Shi2020,MonteroDorta2020B,MonteroDorta2020C,Favole2021,MonteroDorta2021,Engler2021,MonteroDorta2023}).

DM haloes in IllustrisTNG are identified using a friends-of-friends (FOF) algorithm with a linking length of 0.2 times the mean inter-particle separation \citep{Davis1985}. The {\sc subfind} algorithm \citep{Springel2001,Dolag2009} is in turn used to identify subhaloes. Galaxies in this context are defined as subhaloes containing a non-zero stellar mass component. Our redshift evolution analysis employs the merger trees from SubLink \citep{Rodriguez-Gomez2015} in order to select and follow back the main branch of subhaloes at $z=0$. Both the SubLink merger trees and those obtained using 
{\sc LHaloTree} \citep{Springel2005} are publicly available on the TNG database. According to \citealt{Nelson2018}, these two algorithms converge to similar results in a ``population-average sense".

Several subhalo and halo properties from TNG are employed in this work. For haloes, we use the virial mass of the host halo, $M_{\rm host}$ [$h^{-1} {\rm M_{\odot}}$], defined as the total mass enclosed within a sphere of radius $R_{\rm vir}$, where the density equals 200 times the critical density. From $M_{\rm host}$ and the merger trees provided in the database, we compute both the redshift-dependent peak mass, \Mpeakzall, and the half-peak-mass formation time $t_{\rm form}$ (i.e., the time at which half the $z=0$ peak mass \Mpeakz has formed). For the simulated galaxies, we use the stellar mass, $M_\ast{}$ [$h^{-1} {\rm M_{\odot}}$], computed as the sum of the mass of all stellar particles and gas cells bound to each subhalo. Of course, the 3D Cartesian coordinates of galaxies are also used in this analysis. 

Finally, we impose a simple selection based on the analysis of \cite{MonteroDorta2021}. We only analyse haloes above $\log_{10}( M_{\rm host}[h^{-1}{\rm M_\odot}]) = 11$, in order to ensure good resolution in terms of DM particles. For galaxies, we impose a stellar mass cut $\log_{10}( M_{\rm *}[h^{-1}{\rm M_\odot}]) = 8.75$. This cut is not particularly restrictive in terms of resolution, which allows us to maintain a large number of subhaloes that should contain galaxies. Importantly, we are not splitting galaxies by any internal property, so this choice should not bias our results in any significant extent.


\section{The cosmic-web description: DISPERSE}
\label{sec:environment}

This work analyses the relation between secondary bias and the location of galaxies within the cosmic web. The cosmic-web description is based on the Supplementary Catalog of Cosmic Web Distances (SCCWD, 
\cite{Duckworth2020a,Duckworth2020b}, a public dataset obtained by applying the Discrete Persistent Structures Extractor \citep[DisPerSE; ][]{Sousbie2011a,Sousbie2011b} to the TNG300 simulation box. As mentioned above, DisPerSE is a multiscale structure identifier that detects persistent topological features such as peaks, voids, walls and, particularly, filamentary structures. The algorithm draws on discrete Morse theory \citep[see][]{Morse1934}, on the basis that the cosmic web can be described in terms of a mathematical equivalent called Morse complex, i.e., a set of manifolds which, in the cosmological context, can be related to the aforementioned structures. From this object, the critical points of the density field, which correspond to the locations where the field gradient vanishes, can be derived (i.e., the maxima, minima or saddle points of the density field). In order to determine the density field from the input particle distribution, the algorithm uses the Delaunay tessellation field estimator \citep[DTFE; see][]{vandeWeygaert2009}. SCCWD employs a value of $\sigma=4$ for the ``persistence'' parameter, which represents the robustness of the critical point determination. For more information on the procedure and technical aspects, see the DisPerSE webpage \footnote{\url{http://www2.iap.fr/users/sousbie/web/html/indexd41d.html?}}.

Importantly, in the application of DisPerSE on TNG300 performed by \cite{Duckworth2020a} and \cite{Duckworth2020b} (the one that we employ here), the discrete set of points used to estimate the density field are galaxies with a minimum stellar mass of 10$^{8.5}h^{-1} {\rm M_{\odot}}$, which resembles what could be achievable observationally. An interesting aspect to explore in the future is the dependence of our results on the way the density field (and, consequently, the cosmic web) is determined (i.e. if haloes or DM particles are employed). 

As a result of the application of this automatic identification, the SCCWD downloaded from the TNG webpage provides the distances of each galaxy in the box to the 
nearest of each of the following critical points/structures in the density field:

\begin{itemize}
\item Nodes, defined as global maxima. They can be thought of as the centres from which the segments of the cosmic web emerge from or converge into. The distance to nodes is denoted as $d_{\rm node}$ in this work. 
\item Voids, defined as global minima. They represent the empty regions in the network. The distance to voids is denoted as $d_{\rm min}$ in this work. 
\item Saddle points, defined as critical points that are neither minima nor maxima and where one or two dimensions are collapsing (these are called 1D or 2D saddle points, respectively). Within the cosmic web, these points mark regions through which material flows from one node to the other. The distances to 1D and 2D saddle points are denoted as $d_{\rm sadd1}$ and $d_{\rm sadd2}$, respectively. 
\item Filaments, which are 1D structures consisting of a pair of arcs originating from a given saddle point and joining two extrema together. In our context, these structures correspond to the sections of the cosmic web that go from node to node. The catalogue provides the distance to the nearest filament segment, $d_{\rm skel}$ (note that the term ``skeleton'' is employed in this context to represent filaments). 

\end{itemize}

Figure \ref{fig:maps} allows as to visualise the DisPerSE distance parameters across the cosmic web in a slice of 30 x 205 X 205 $h^{-3}$Mpc$^3$. Note that the values of $\{d_i\}$, indicated by the colour code,  do not represent, for a given galaxy, its local environment directly, but the distance to the specific cosmic-web features described above. It is easy to see, that low values of $d_{\rm node}$ correspond, predominantly, to high-density regions, whereas low values of $d_{\rm skel}$ trace the filamentary skeleton-like network. Low values of $d_{\rm min}$, conversely, mark the low-density areas (voids), although these are visually affected by projection effects. Finally, saddle points are mainly found inside filaments, which explains that the low values of $d_{\rm sadd1}$ and $d_{\rm sadd2}$ also trace these structures. 
 
\begin{figure*}
	\includegraphics[width=2.0\columnwidth]{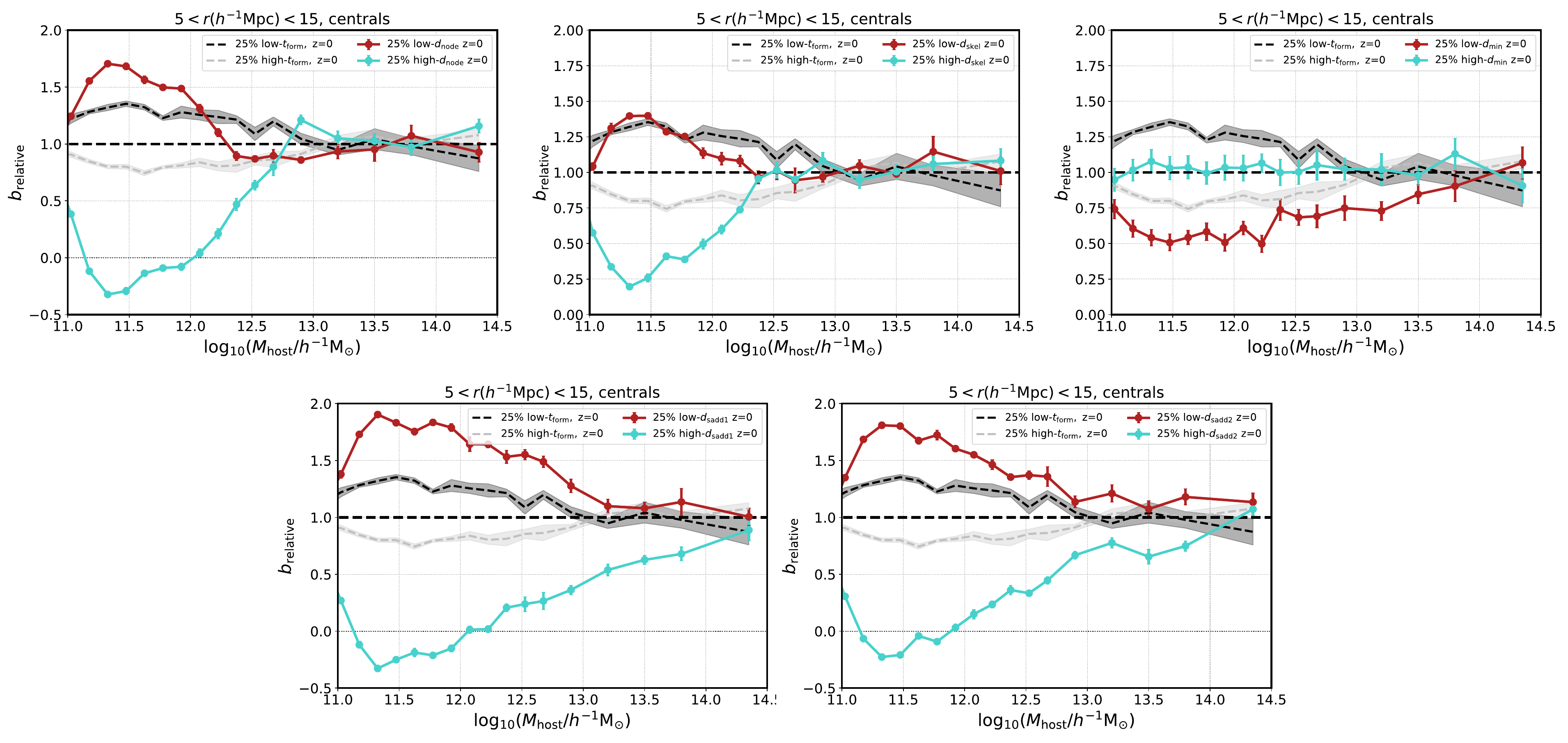}
    \caption{Secondary bias measurement based on the DisPerSE cosmic-web distance parameters as compared to the formation time ($t_{\rm form}$) dependence (also known as assembly bias). In the upper row, from left to right, the secondary bias obtained after splitting the central galaxy population by distance to nodes ($d_{\rm node}$), filaments ($d_{\rm skel}$), and voids ($d_{\rm min}$) is presented. In the lower row, distances to 1D saddle points ($d_{\rm sadd1}$) and 2D saddle points ($d_{\rm sadd2}$) are employed to perform the analysis. In each panel, the solid lines correspond to the relative bias measured for the 25$\%$ higher/lower-distance populations (dark red/cyan), whereas the dashed lines correspond to the assembly bias trend. Results for low formation time (older haloes) are represented in black, and those for high formation time (younger haloes), in grey. Errors for all measurements are obtained using a jackknife technique.}  
    \label{fig:AB1}
\end{figure*}

\section{The dependence of assembly bias on the cosmic web}
\label{sec:ABCW}

\subsection{Definitions and assumptions}
\label{sec:definitions}

In the context of {\it{assembly bias}}, multiple effects are often named by different authors using the same terminology. For this reason, it is convenient to establish our assumptions and definitions clearly to avoid confusion. We, therefore, define here: 

\begin{itemize}
\item Secondary halo bias/secondary galaxy bias, as the secondary dependencies of halo/galaxy bias on either internal halo properties or environmental properties, at fixed halo mass. Note that for secondary galaxy bias, the dependence of galaxy clustering must be connected to a halo or environmental property as well.  
\item Halo assembly bias, as the specific secondary halo bias dependence on formation time (or the accretion history of haloes, more generally).
\item Galaxy assembly bias, as the secondary galaxy bias dependence on formation time (or the accretion history of haloes, more generally). Note that alternative definitions are employed in the literature, such as those based on the occupancy variations across haloes. We adopt the simplest definition where galaxy assembly bias is just a direct manifestation of halo assembly bias on the galaxy population. 

Since we use central galaxies and we are not splitting by any galaxy properties, the galaxy effects are nothing but the result of imposing a particular selection on the halo population. For this reason, the halo/galaxy term is often dropped throughout this work for simplicity. Note also that we are explicitly extending the definition of secondary bias to external properties, which is not always the case in the previous works. 

\end{itemize}

\subsection{Secondary bias and the DisPerSE distances}
\label{sec:AB}

The DisPerSE parameters provide a detailed characterisation of the cosmic web in terms of a set of physically-motivated critical density points, $\{d_i\}$. In this section, we use these parameters to investigate the connection between secondary bias and the cosmic web. To this end, clustering is measured for galaxy subsets split by distance to these cosmic-web features at fixed halo mass. Although these distances are not internal halo properties, they can provide important insights into the nature and origins of assembly bias. 

In order to measure secondary bias, we follow a common procedure based on the {\it{relative bias}}, $b_{\rm relative}$, between conveniently chosen subsets of galaxies (e.g., \citealt{2018Salcedo, SatoPolito2019}). We measure the (3D) real-space two-point correlation function, $\xi(r)$, using the Landay-Szalay estimator \citep{Landy1993}. Following \cite{MonteroDorta2020B}, the relative bias between galaxy subsets is subsequently determined on the basis of the cross-correlation between subsets and the entire sample, which maximises the signal-to-noise of the measurement (this is important given the small volume of TNG300). For a given halo-mass bin, $M_{i}$, and subset $\mathcal{S}$, the relative bias can be measured as: 

\begin{equation}
   b_{\rm relative}(r,\mathcal{S}|M_i) = \frac{\xi_{[\mathcal{S},{\rm all}]}(r)}{\xi_{[M_{i},{\rm all}]}(r)},
   \label{eq:relative_bias}
\end{equation}
 
\noindent where $\xi_{[S,{\rm all}]}$ is the cross-correlation between all objects in the subset and all objects in the sample, and $\xi_{[ M_i,{\rm all}]}$ is the cross-correlation between all objects in the halo-mass bin and the entire sample as well. Here, the subsets $\mathcal{S}$ are defined based on the DisPerSE distances in terms of a certain percentile (either quartiles or the median of the distribution). The computation of errors is based on a standard jackknife technique, where the TNG300 box is divided into 8 sub-boxes ($ L_{{\rm sub-box}} = L_{{\rm box}}/2 = 102.5$ $h^{-1}$Mpc). The relative bias is averaged over scales 5-15 $h^{-1}$Mpc. 

Eq.~\ref{eq:relative_bias} is employed in Fig.~\ref{fig:AB1}, which displays the secondary dependence of galaxy clustering on the different DisPerSE parameters described in Section \ref{sec:environment}, along with the assembly bias signal, i.e., the dependence of galaxy clustering on the formation time of the halo at fixed halo mass. As mentioned before, we use the half-mass formation time as our proxy for halo age; for discussion on alternative ways of defining this property in the context of assembly bias see \citealt{Li2009, MonteroDorta2021}. We emphasise that the analysis is restricted to the halo mass range $\log_{10}( M_{\rm host}[h^{-1}{\rm M_\odot}]) > 11$ in order to avoid resolution problems for lower mass haloes (and their corresponding central galaxies, see \citealt{MonteroDorta2020B}). As in Fig. \ref{fig:maps}, the upper panels present, from left to right, the measurements for $d_{\rm node}$, $d_{\rm skel}$, and $d_{\rm min}$, respectively, whereas $d_{\rm sadd1}$ and $d_{\rm sadd2}$ are shown in the bottom row. All subsets employed in this figure are defined in terms of quartiles (25$\%$ higher/smaller values for the DisPerSE distances and half-mass formation time).

\begin{figure*}
	\includegraphics[width=1.5\columnwidth]{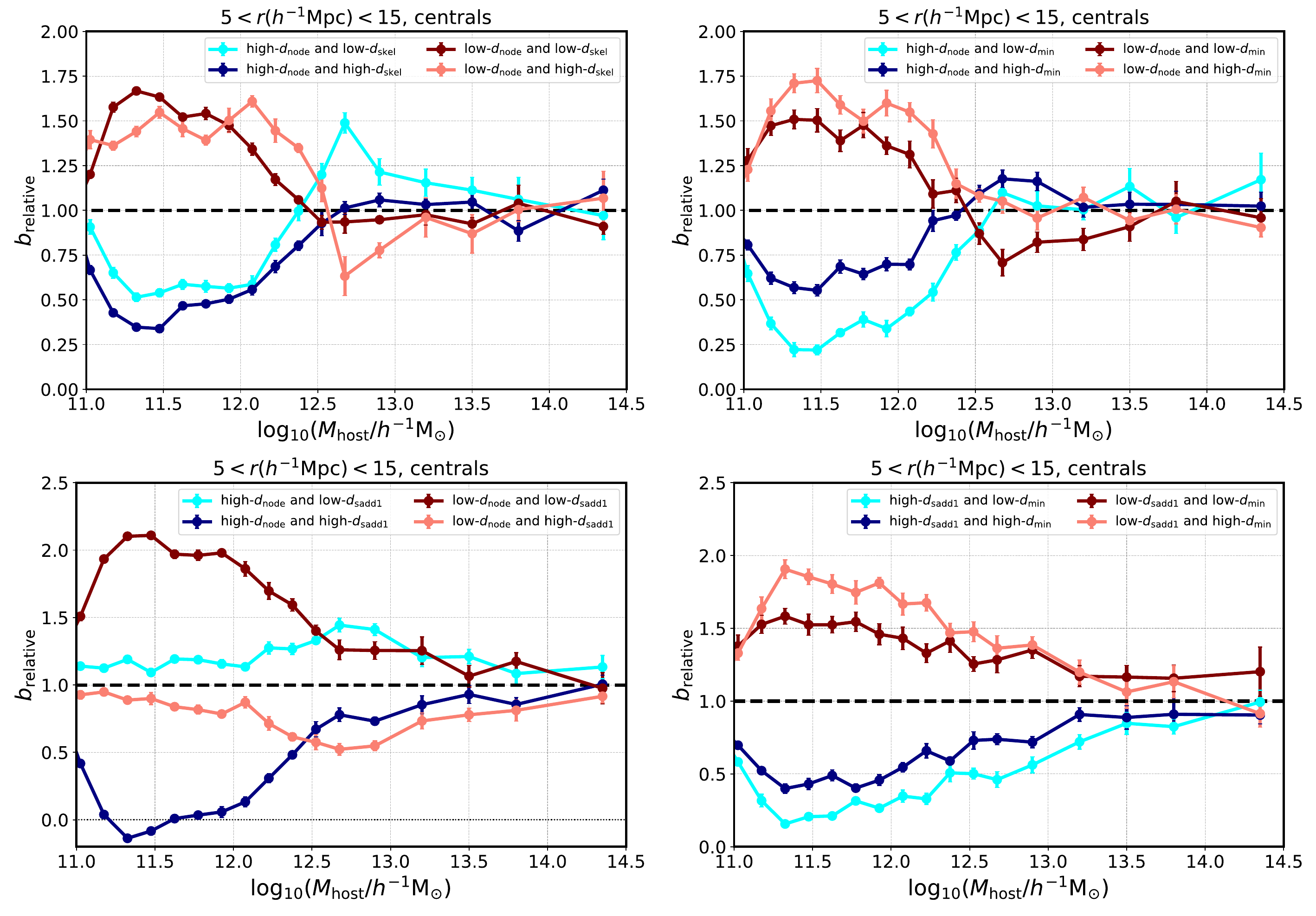}
    \caption{Secondary bias measurements using subsets defined in terms of pairs of DisPerSE cosmic-web distances, in a similar format to that of Fig. \ \ref{fig:AB1}. In the upper row, from left to right, 50$\%$ subsets (i.e. using the median as a demarcation) in $d_{\rm node}$-$d_{\rm skel}$ and $d_{\rm node}$-$d_{\rm min}$ are employed, respectively. The lower panels show results for  $d_{\rm node}$-$d_{\rm skel}$ (left) and $d_{\rm node}$-$d_{\rm min}$ (right). Errors for all measurements are obtained using a jackknife technique.}
    \label{fig:AB2}
\end{figure*}

The first thing to notice from Fig.~\ref{fig:AB1} is that splitting the $z=0$ central galaxy population using the DisPerSE cosmic-web indicators produces varying degrees of secondary bias with different mass dependencies. These signals generally exceed the TNG300 assembly bias prediction by a large margin, particularly towards the low-mass end. In the case of $d_{\rm node}$, Fig.~\ref{fig:AB1} shows that, at fixed halo mass, central galaxies in haloes close to nodes are more tightly clustered than those far from them at $\log_{10}( M_{\rm host}[h^{-1}{\rm M_\odot}]) \lesssim 12.5$. In this range, the difference in the relative bias with respect to the subset of galaxies that lie far from nodes reaches a value of $\sim 2$. This secondary bias effect is significantly stronger than assembly bias, which only reaches a differential signal of around $\lesssim 0.5$ at the low-mass end. Qualitatively, the signal produced by $d_{\rm skel}$, the distance to filaments (middle panel, upper row), is similar, although with a significantly smaller amplitude.

When galaxies are split by distance to a void (i.e., density minimum, $d_{\rm min}$), a statistically-significant secondary bias signal is also measured, albeit with a qualitatively different shape as compared to the previous parameters. In this case, central galaxies far from voids display no secondary bias (i.e., their clustering at fixed halo mass is indistinguishable from that of the entire population). Galaxies near voids, conversely, are less clustered, with the combined amplitude of the secondary bias signal remaining similar to that of formation time (assembly bias). 

Finally, the distance to the saddle points (including the 1D and the 2D versions, lower row in Fig.~\ref{fig:AB1}) also produces interesting secondary bias signals, with galaxies closer to these critical points being more highly biased. For $d_{\rm sadd1}$, we find the strongest signal among all parameters, with a differential amplitude well above a value of 2. In fact, a statistically significant effect is measured even at the highest masses, unlike what is known for formation time. Note that significant levels of high-mass secondary bias have been measured for other halo properties, most notable concentration and spin, see, e.g. \citealt{Gao2007, faltenbacher2010, 2018Salcedo, SatoPolito2019}). The 2D saddle point yields a similar trend, although the amplitude of the signal decreases slightly. Importantly, when compared to the results of \cite{MonteroDorta2020B}, the signal measured for $d_{\rm sadd1}$ exceeds significantly that of any internal halo property measured from TNG300. These results highlight the prominent role of saddles as reference points in the structure of the cosmic web, as previously pointed out by several authors (see, e.g., \citealt{Codis2015, Musso2018})

Fig.~\ref{fig:AB1} demonstrates that the cosmic-web environment has, as expected, a significant impact on the clustering of haloes and galaxies, even when the measurement is performed at fixed halo mass. In Fig.~\ref{fig:AB2}, this connection is analysed in more detail by splitting the galaxy sample by pairs of cosmic-web distance indicators. Using the same format of Fig.~\ref{fig:AB1}, we show in Fig.~\ref{fig:AB2} the secondary bias trends for $50\%$-subsets in the planes of $d_{\rm node}$-$d_{\rm skel}$ (upper left), $d_{\rm node}$-$d_{\rm min}$ (upper right), $d_{\rm node}$-$d_{\rm sadd1}$ (lower left), and $d_{\rm sadd1}$-$d_{\rm min}$ (lower right). Hereafter, we will only analyse 1D saddle pints for simplicity, since these are the ones that display the strongest signal. 

At the low-mass end, the distances to nodes and filaments from DisPerSE are correlated (Pearson correlation coefficient, PCC, of 0.6-0.7). This explains why, qualitatively, galaxies at fixed $d_{\rm node}$ or $d_{\rm skel}$ displays similar levels of secondary bias. At least, there is no clear trend in the upper left panel of Fig.~\ref{fig:AB2}. At a reference halo mass of $\log_{10}( M_{\rm host}[h^{-1}{\rm M_\odot}]) \simeq 11.5$, the maximum separation in relative bias is measured between galaxies closer to nodes and filaments, and galaxies far from these critical points. The situation is different when subsets based on $d_{\rm node}$ and $d_{\rm min}$ are simultaneously analysed (upper right panel); these two parameters are not significantly correlated. At fixed halo mass, central galaxies close to nodes are clearly more clustered if they happen to be also far from voids, whereas those that are far from nodes are even less clustered if they are near voids. 

In the lower panels of Fig.~\ref{fig:AB2} a similar exercise is performed using the distance to the 1D saddle points. On the left-hand side, the maximum separation is obtained between galaxies close to both nodes and saddle points and those far from them (PCC of $\sim 0.4$ for $\{d_{\rm node}, d_{\rm sadd1}\}$). For galaxies close to nodes and far from saddle points, and vice versa, the secondary bias signal inverts, with respect to what it would be expected for $d_{\rm node}$. This is not surprising, since the magnitude of the signal, as shown in Fig.~\ref{fig:AB1} is stronger for $d_{\rm sadd1}$. Finally, the results for the lower right panel are again consistent with the fact that $d_{\rm min}$ is uncorrelated with $d_{\rm sadd1}$: there is a clear trend indicating that the closer to saddles and farther from voids the higher the clustering.

It is interesting to note from Fig.~\ref{fig:AB2} that there are certain combinations of distance parameters for which significant crossovers in the signals are observed. In particular, clear inversions are found at $\log_{10}( M_{\rm host}[h^{-1}{\rm M_\odot}]) \sim 12.5$ for galaxies far from nodes and close to filaments, and also for the "opposite" population (close to nodes and far from filaments, upper left panel). Something similar, albeit less prominent, is observed for objects close to nodes and voids, and for those far from both these critical points as well (upper right panel). These crossovers could potentially be related to similar features measured previously for internal halo properties such as concentration, although in that case, the inversion is found at $\log_{10}( M_{\rm host}[h^{-1}{\rm M_\odot}]) \sim 13$ (e.g., \citealt{Gao2007, Wechsler2006, SatoPolito2019}). Note that a less obvious inversion was already found for $d_{\rm node}$ in Fig.~\ref{fig:AB1}, which suggests that this feature arises from a mechanism that acts around nodes and is amplified when an additional condition is imposed. 

\begin{figure*}
	\includegraphics[width=2\columnwidth]{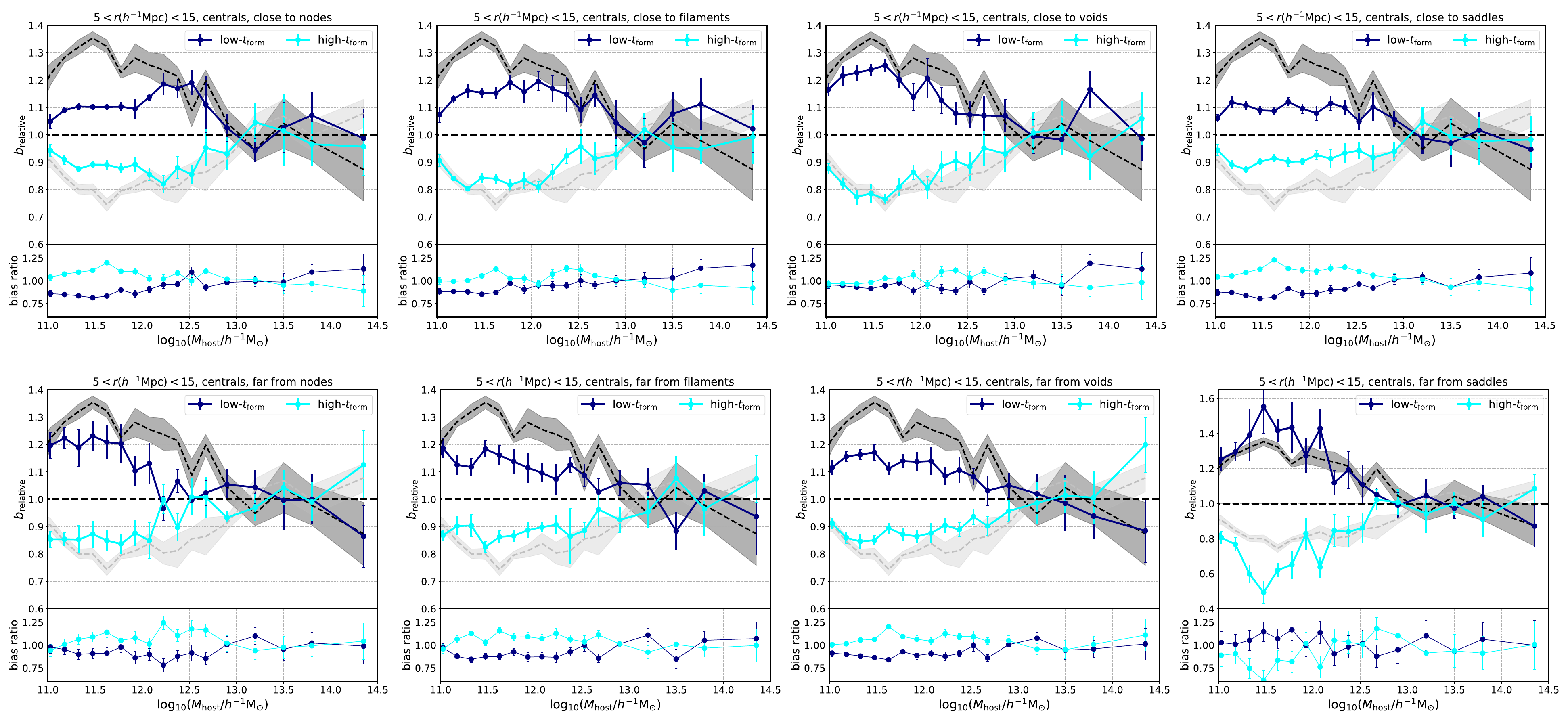}
    \caption{Assembly bias measurements (i.e., dependence of clustering on formation time, $t_{\rm form}$, at fixed halo mass) for galaxies close (upper panels) and far (lower panels) from nodes ($d_{\rm node}$), filaments ($d_{\rm skel}$), voids ($d_{\rm voids}$), and saddle points ($d_{\rm sadd1}$). This figure uses a similar format to that of Fig. \ref{fig:AB2}, but in this case, 50$\%$ subsets based on both $t_{\rm form}$ and the DisPerSE parameters are employed (i.e. with the median as the demarcation). In each panel, both the global assembly bias signal (for the entire population, black/grey dashed lines) and the assembly bias signal conditioned to distance to critical points (blue/cyan solid lines) are displayed for comparison. The sub-panels show the ratio between the global signal and the conditioned signal, for the high-$t_{\rm form}$ (dark blue) and low-$t_{\rm form}$ (cyan) subsets, where errors have been computed propagating the uncertainties on the measurements. }
    \label{fig:AB3}
\end{figure*}

\begin{figure}
	\includegraphics[width=\columnwidth]{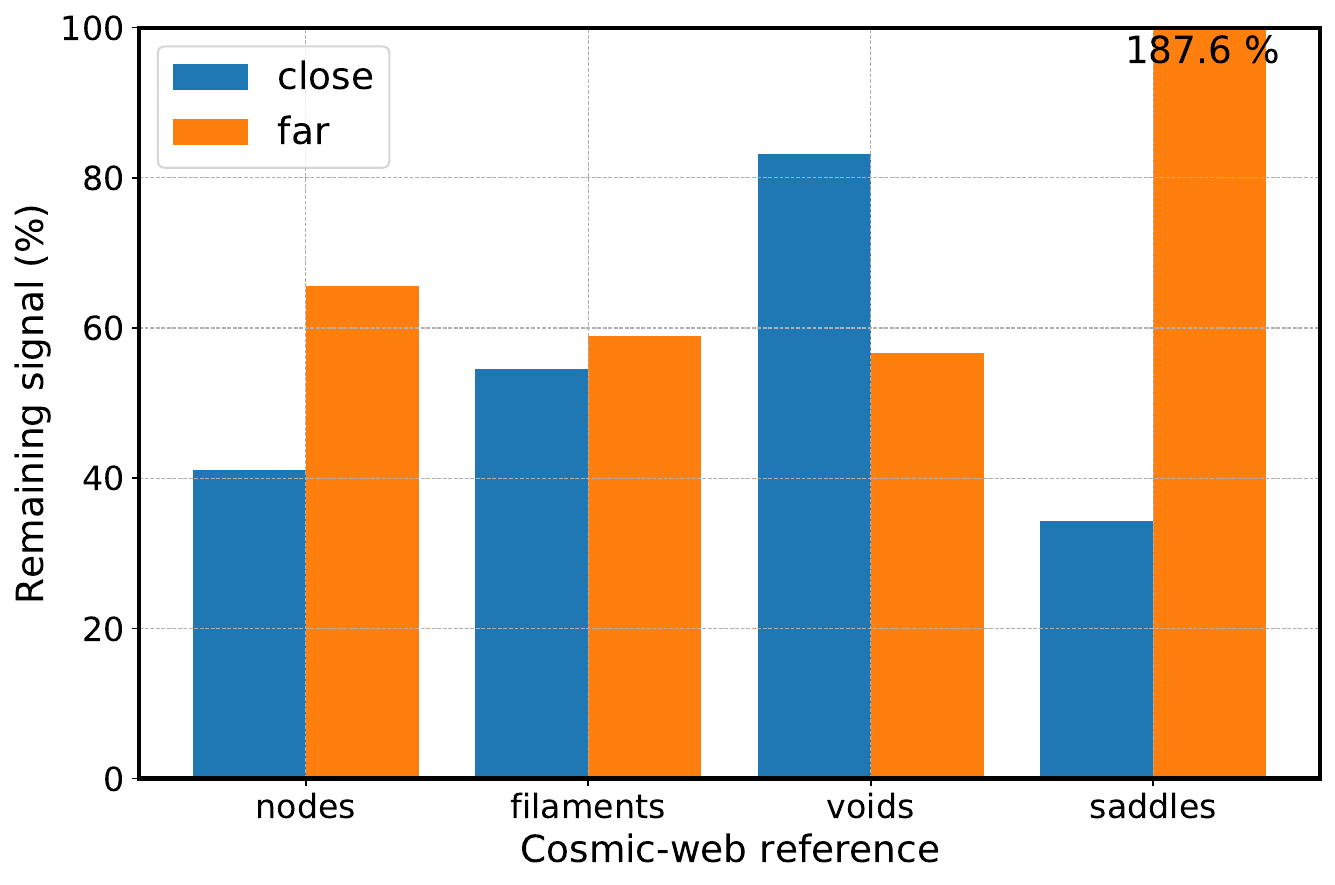}
    \caption{The percentage of low-mass assembly bias at $\log_{10}( M_{\rm host}[h^{-1}{\rm M_\odot}]) \simeq 11.5$ that remains after conditioning the secondary bias measurement to objects close to or far from nodes, filaments, voids, and saddles. This chosen halo mass approximately corresponds to the mass at which the maximum assembly bias signal is measured. The remaining fraction is obtained by dividing the differential assembly bias signal obtained for galaxies in a certain cosmic-web environment and that measured for the entire population (see text). Note that, for the sake of clarity, the y-axis range has been limited to 100$\%$, despite the fact the remaining fraction for objects far from the saddles reaches a value of 187.6 $\%$ (note that this implies that the conditioned signal surpasses the global signal).  }
    \label{fig:percentage}
\end{figure}

The fact that conditioning the clustering measurement based on the distance to conspicuous points in the cosmic web produces variations in bias at the secondary level is, to some extent, not surprising (i.e., for lack of a better word, we are ``biasing" our correlation function measurement on purpose when splitting the sample at fixed halo mass). However, both the particular amplitude and the shape of the dependencies, which change as a function of halo mass and distance parameters, are relevant in terms of characterising the large-scale structure. It is also relevant that some of these dependencies are qualitatively similar to the assembly bias trend. Important for this work, the connection with assembly bias that we explore in more detail in the following section, can shed light onto the physical origins of the signal.

\subsection{The connection with formation time}
\label{sec:formation_time}

\begin{figure*}	
\includegraphics[width=2\columnwidth]{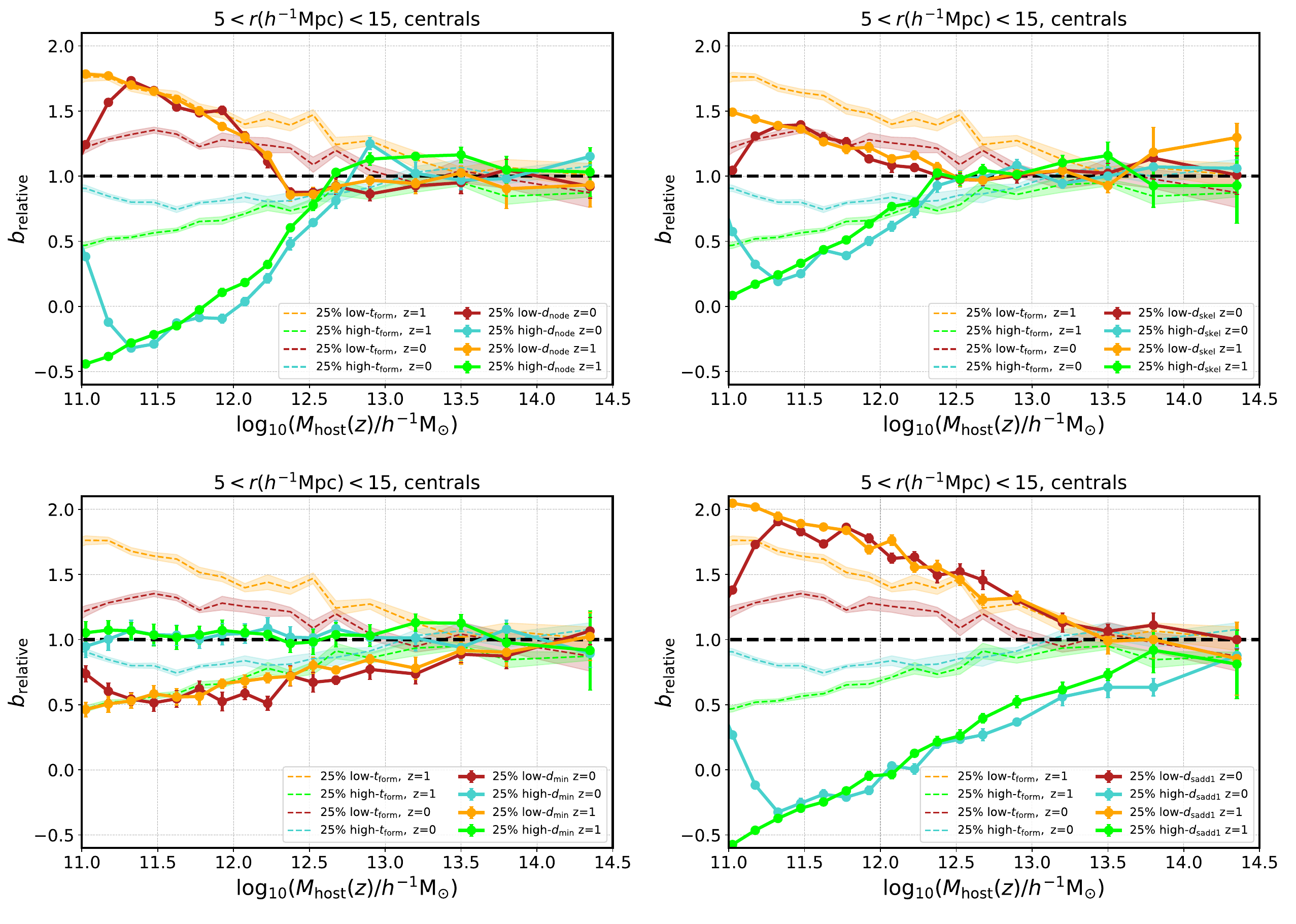}
   \caption{The evolution of central galaxy secondary bias    
   based on the distance to nodes ($d_{\rm node}$, upper left panel), filaments ($d_{\rm skel}$, upper right), voids ($d_{\rm voids}$, lower left), and saddle points ($d_{\rm sadd1}$, lower right), using a similar format to that of Fig. \ref{fig:AB1}. Here, the evolution is measured along the merger tree, i.e., for the progenitors of $z=0$ galaxies.
    In each panel, solid lines represent results based on the cosmic-web distances, whereas dashed lines display the assembly bias ($t_{\rm form}$) signal. Errors for all measurements are obtained using a jackknife technique.}
    \label{fig:AB_z}
\end{figure*}

As discussed in the Introduction, effects related to the location of halos in different cosmic web environments have been previously invoked to provide a physical explanation for low-mass assembly bias (e.g., \citealt{Dalal2008,Borzyszkowski2017,Musso2018}). As described in detail in \cite{Borzyszkowski2017}, the signal might arise from a population of ``stalled" haloes in filaments whose growth is halted early on, as opposed to ``accreting" haloes in nodes which display a more continuous accretion (note that for assembly bias to occur, both populations must have the same mass). In order to investigate this hypothesis, we have performed a similar analysis to that of Fig.~\ref{fig:AB2} but combining cosmic-web information with the formation time of haloes in our selections. In the upper panels of Fig.~\ref{fig:AB3}, the assembly bias signal is measured for galaxies close, from left to right, to nodes, filaments, voids, and saddle points, whereas the lower panels display the measurement for galaxies far from these critical points. Here, as in Fig.~\ref{fig:AB2}, 50$\%$ subsets are employed in order to maintain adequate statistics for the correlation function determination. Note that ``close" here means something different for each cosmic-web indicator (the same happens in Fig.~\ref{fig:AB2}). The 50$\%$ demarcation corresponds to distances to nodes of the order of 3-4 $h^{-1}$Mpc for haloes below  $\log_{10}( M_{\rm host}[h^{-1}{\rm M_\odot}]) \gtrsim 12$, dropping to smaller distances for higher masses (for the most massive haloes it is of the order of a dozen $h^{-1}$kpc). Something similar happens for $d_{\rm skel}$, although in this case even at small halo masses the 50$\%$-limit distance remains within 1-2 $h^{-1}$Mpc. Conversely, for both $d_{\rm min}$ and $d_{\rm sadd1}$ the demarcation is found at higher distances, and almost independently of halo mass: 5-6 $h^{-1}$Mpc for the distance to filaments and as far as $\sim$30 $h^{-1}$Mpc for voids.

Fig.~\ref{fig:AB3} clearly shows that: 1) the assembly bias signal tends to decrease in amplitude when galaxies are restricted to some particular cosmic environments, 2) the signal, does not quite vanish at the low-mass end, which might indicate that the assembly bias trend cannot be totally explained by the different cosmic-web environments. The attenuation of the assembly bias signal depends also on the halo mass range. Close to nodes, it seems fairly unperturbed at $\log_{10}( M_{\rm host}[h^{-1}{\rm M_\odot}]) \gtrsim 12.5$, to the extent that the uncertainties in the measurement allow. Below this halo mass, the assembly bias signal is reduced to (a difference in the relative bias of) 0.2, approximately. A similar result is obtained for halos close to saddle points, but the attenuation starts at $\log_{10}( M_{\rm host}[h^{-1}{\rm M_\odot}]) \sim 13$. These two conditions are the ones that produce the largest reduction in the assembly bias signal.  

Galaxies close to filaments and voids maintain a higher level of assembly bias, particularly the latter. In both cases, the clustering of the low-bias subset remains almost the same, whereas the relative biases of the more clustered galaxies are reduced by $\sim 50\%$ for $d_{\rm skel}$ and $\sim 30\%$ for $d_{\rm min}$ (these reduction factors are computed with respect to the $b_{\rm relative}=1$ demarcation). It seems, therefore, that the effect of the environment is weak near voids, thus allowing us to recover a more pristine assembly bias signal.

Fig.~\ref{fig:AB3} also displays results for haloes far from any critical point (lower row). Again, results depend on the estimator and halo mass range considered. For $d_{\rm node}$, the attenuation of the assembly bias signal is less significant at small masses but is stronger at intermediate masses (i.e., around $\log_{10}( M_{\rm host}[h^{-1}{\rm M_\odot}]) \sim 12.5$), as compared to the close-distance results. For $d_{\rm skel}$, the signal is actually slightly more attenuated when the large-distance condition is imposed. That is clearly also the case for $d_{\rm min}$: haloes far from voids tend to display a lower assembly bias signal, which is consistent with the fact that they are also closer to cosmic-web structures. Finally, when haloes far for saddle points are analysed, the assembly bias signal actually grows in some mass ranges, as compared to the general, underlying trend. 

The description of the cosmic web in terms of a set of critical points in the density field is highly complex, which makes it non-trivial to extract a comprehensive picture from Figs.~\ref{fig:AB3}. In order to illustrate more clearly the effect of conditioning the assembly bias measurement to particular environments, in Fig.~\ref{fig:percentage} we pick a characteristic halo mass, at $\log_{10}( M_{\rm host}[h^{-1}{\rm M_\odot}]) \simeq 11.5$, where the maximal assembly bias signal is found, and show the fraction of this signal that remains for each distance estimator (close and far, as in Fig.~\ref{fig:AB3}). This fraction is obtained by taking the ratios of the differential signals (subtracting the high and low-$t_{\rm form}$ relative bias values) for the conditioned and the global configurations (in absolute value). The results displayed in Fig.~\ref{fig:percentage} emphasise again the role of saddle points for secondary bias, as they display, by far, the maximum split between the far and close conditions. They also reinforce the idea that close to voids the secondary dependence of galaxy bias on the formation history of haloes appears to remain fairly unperturbed (more than 80$\%$ of the signal remains).

\subsection{Redshift evolution}
\label{sec:z_evolution}

The evolution of secondary bias (and in particular, assembly bias) has been addressed in several previous works (e.g., \citealt{Wechsler2006, faltenbacher2010,Contreras2019,Tucci2021}). From these and other related works, a consensus has been reached that the redshift dependence of the signal scales, at least to a first approximation, with the peak height of fluctuations, $\nu = \delta_c(z)/\sigma(M_{\rm host},z)$, where $\delta_c(z)$ is the redshift-dependent density contrast for collapse and $\sigma(M_{\rm host},z)$ is the rms of the linear overdensity field on a sphere containing a mass of  $M_{\rm host}$ at redshift $z$. This explains why the signal, when analysed across independent snapshots, decreases with redshift at fixed halo mass, as progressively smaller $\nu$ are mapped for the same halo mass (e.g., \citealt{Wechsler2006, faltenbacher2010,Contreras2019, Tucci2021}). A different way to look at the evolution of secondary bias is ``along the merger tree", that is, only for the progenitors of the $z=0$ population. \cite{MonteroDorta2021} showed that, at fixed $z=0$ peak mass, the assembly bias signal actually increases with redshift when only the progenitors of the $z=0$ population are considered, and the formation time is computed at $z=0$. As discussed in \cite{MonteroDorta2021} this trend remains, although it weakens significantly, when the instantaneous host mass, $M_{\rm host}(z)$ is employed.

We took this approach to measure the evolution of the secondary bias emerging from the DisPerSE cosmic-web parameters. In the same format of previous figures, Fig.~\ref{fig:AB_z} shows the secondary bias signal for $d_{\rm node}$, $d_{\rm skel}$, $d_{\rm min}$, and $d_{\rm sadd1}$ for central galaxies at $z=0$ (the same trends shown bin Fig. \ref{fig:AB1}, red/blue solid lines) and for their progenitors at $z=1$ (as long as they are centrals; orange/magenta solid lines). Here, the halo mass corresponds to the instantaneous mass at $z=0$ and $z=1$ (note that a similar analysis could be performed for peak mass). These measurements are compared with the evolution of the assembly bias signal within the same redshift range (dashed lines). As shown in Fig.~\ref{fig:AB_z}, the assembly bias signal for the progenitors of $z=0$ galaxies increases with redshift in a fairly progressive way (as shown in \citealt{MonteroDorta2021}). Note, again, that the $z=0$ formation time is used to split galaxies at $z=1$.

The most remarkable result from Fig.~\ref{fig:AB_z} is the almost complete lack of evolution (beyond the level of noise of the measurement) for the secondary bias produced by the DisPerSE cosmic-web parameters. The central galaxies that are progenitors of the $z=0$ population display almost the exact same trends as their low-redshift counterparts for $d_{\rm node}$, $d_{\rm skel}$, $d_{\rm min}$, and $d_{\rm sadd1}$, despite the evolution of clustering and the growth of structure. The most notable difference is visible at the low-mass end, where all the $z=0$ trends tend to converge to 1, whereas the $z=1$ lines keep separating from each other. We attribute this effect to resolution, since a similar drop is measured for $t_{\rm form}$ (see also \citealt{MonteroDorta2020B,MonteroDorta2021}). 

We have also checked that the conclusions presented in Figs. \ref{fig:AB3}-\ref{fig:percentage} regarding the attenuation of assembly bias in different environments at $z=0$ remain almost unchanged when the analysis is performed at $z=1$. This implies that the connection between assembly bias and the cosmic web, as presented in this work, is already in place at $z=1$.

\subsection{Correlation analysis}
\label{sec:correlation}

\begin{figure}	
    \centering
    \includegraphics[width=0.7\columnwidth]{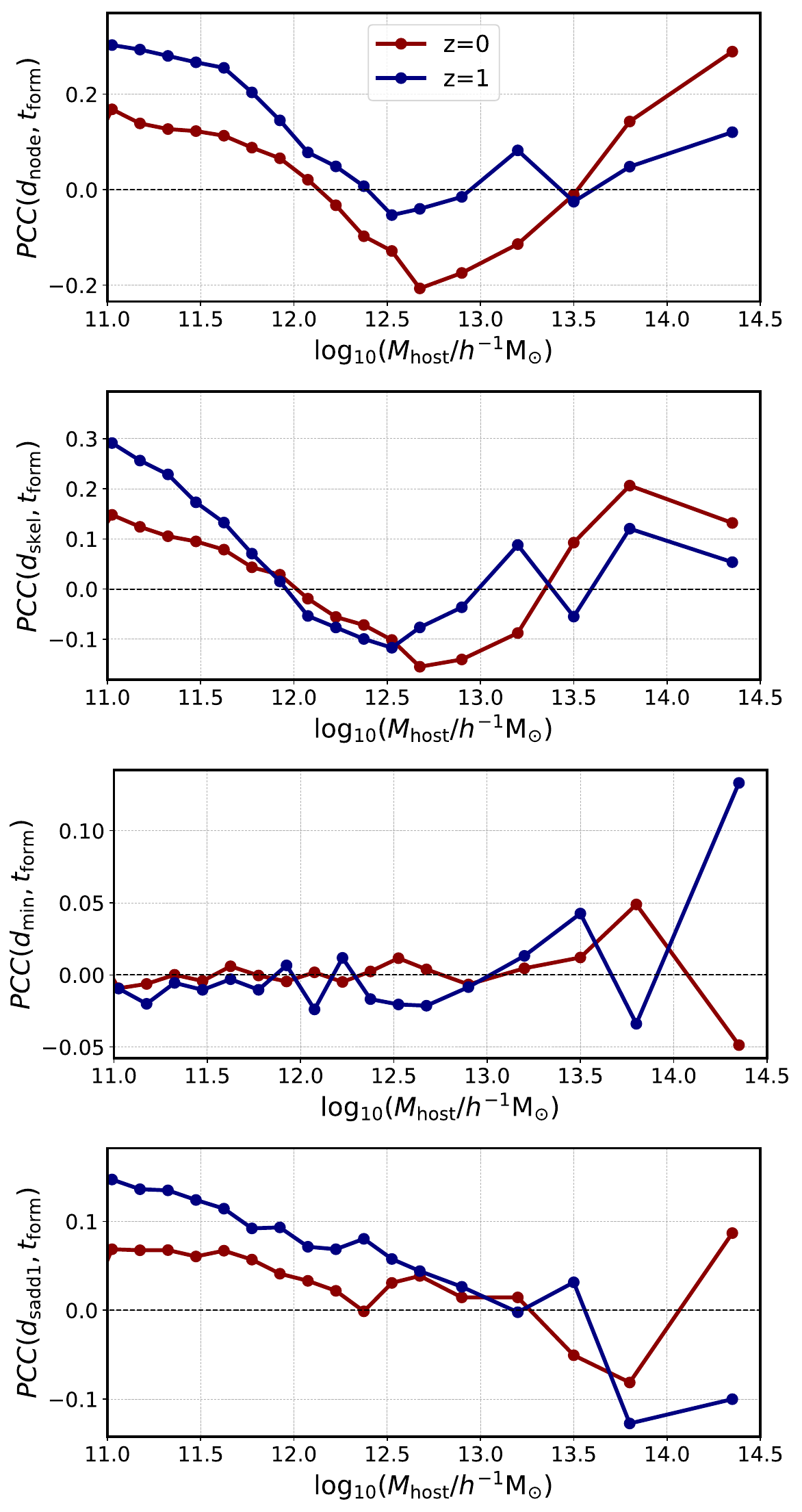}
    \caption{The Pearson correlation coefficients (PCCs) between the cosmic-web distance parameters and the formation time of haloes, $t_{\rm form}$, for redshifts $z=0$ and $z=1$.}
    \label{fig:corr}
\end{figure}

The correlation between the distance parameters and formation time for the two redshift snapshots considered is analysed in more detail in Fig.~\ref{fig:corr}, which displays the Pearson correlation coefficients (PCCs) for pairs of properties $\{t_{\rm form}, d_i\}$. This figure shows positive correlations, albeit modest, between $t_{\rm form}$ and the distance to nodes, filaments and saddles, for central galaxies in host haloes of $\log_{10}( M_{\rm host}[h^{-1}{\rm M_\odot}]) \lesssim 12-12.5, 12, 13.5$, respectively. The PCC values are higher in these ranges for $d_{\rm node}$ and $d_{\rm skel}$  ($\sim 0.15-0.3$), as compared to $d_{\rm sadd1}$ ($\sim 0.07-0.16$). There is, therefore, a small but significant tendency for early-formed low-mass haloes to live close to these structures. Above the aforementioned masses, $t_{\rm form}$ becomes anti-correlated with $d_{\rm node}$ and $d_{\rm skel}$ up to $\log_{10}( M_{\rm host}[h^{-1}{\rm M_\odot}]) \lesssim 13.5$. The behaviour for the most massive host haloes becomes erratic, which might be due to the rather poor statistics that we have; not also that assembly bias is very weak at that range. Fig.~\ref{fig:corr} shows no correlation between the distance to voids and formation time\footnote{For simplicity, we have opted not to show the PCCs between the distance parameters themselves. As a summary, at $z=0$, $d_{\rm node}$ and $d_{\rm skel}$ shows high positive correlation at the low-mass end (PCC$\sim 0.7-0.5$, decreasing with mass). The correlation between these parameters and $d_{\rm sadd1}$ is smaller (PCC$\sim 0.3-0.4$ for low-mass haloes) and, again, little correlation is found with respect to $d_{\rm min}$.} (within the mass range of interest).

Fig.~\ref{fig:AB_z} demonstrates that the secondary bias trends emerging from the distance to critical points in the cosmic web is similar at $z=1$. Fig.~\ref{fig:corr} shows, however, that the correlation between the distance parameters and formation time tend to increase with redshift for $d_{\rm node}$, $d_{\rm skel}$, and $d_{\rm sadd1}$ (particularly at the low-mass end, where the secondary bias signals are larger). The interpretation of this result is unclear, so further investigation will be needed to provide more insight on this particular aspect. Note, also, that this result might change if the formation time is computed with respect to $z=1$.

At face value, these PCC results would appear to be in slight contradiction with the simple hypothesis that assembly bias emerges from the existence of populations of same-mass haloes that live inside filaments and very close to nodes (where the latter would tend to be younger). According to Fig.~\ref{fig:corr}, when selecting older haloes (high $t_{\rm form}$), there is a slightly higher preference of selecting haloes close to nodes than close to filaments, than finally close to saddles. However, the differences in the correlation coefficients are rather small. Also, it is rather likely that a simple PCC analysis like this is not capable of capturing the intricacies of the geometry of tides in the cosmic web.

\section{Discussion and conclusions}
\label{sec:discussion}

In this paper, we investigate the connection between the cosmic-web environment and the effect that we call galaxy secondary bias, which is defined here as the secondary dependencies of central galaxy clustering on halo properties (either internal or environmental) at fixed halo mass. To carry out this investigation, we have used the public data resulting from the application of the Discrete Persistent Structures Extractor  \citep[DisPerSE; ][]{Sousbie2011a,Sousbie2011b} to the TNG300 hydrodynamical simulation box. The DisPerSE cosmic-web description is the product of a mathematical analysis of the density field from which a set of characteristic critical points is derived. These critical points in the density field can then be associated with large-scale environments and structures. Nodes and voids in this context correspond to the global maxima and minima of the field, respectively, whereas filaments are viewed as regions of intermediate density that join nodes and pass through so-called saddle points. These saddles are neither global minima nor maxima, but places where the concavity of the density field changes. 

The connection between environment and secondary halo bias has been addressed in the literature from multiple but complementary angles. \cite{Dalal2008} introduced the idea that low-mass halo assembly bias, as opposed to its high-mass counterpart, may be due to a subpopulation of haloes whose accretion has ceased. Subsequently, \cite{Hahn2009} found that an important driver of suppressed growth, by accretion and mergers, is tidal effects dominated by a neighbouring massive halo. \cite{Borzyszkowski2017} suggested that low-mass assembly bias is  associated with the existence of same-mass subpopulations of ``stalled" and ``accreting" haloes typically living, respectively, in filaments and nodes. The impact that the geometry of the tides has on the assembly bias effects is analysed and modelled in \cite{Musso2018}. Using DM-only simulations and also TNG, the secondary halo bias trends have been shown to correlate with the anisotropy of the tidal tensor in, e.g., \citealt{Paranjape2018, Ramakrishnan2019}. Finally, to name some of the relevant works, tidal effects affecting splashback haloes are known to be involved in the inversion of the halo spin bias signal at the low-mass end \citep{Tucci2021}.

The importance of environment for secondary bias that previous works pointed out is clearly reflected on our results based on the DisPerSE topological description, which in fact add a different but related angle to the discussion. At fixed halo mass, the distances to the critical points associated with nodes, filaments, voids and saddles display significant secondary bias signals, which often exceed those measured for internal halo properties by a significant amount. We have also shown that restricting the analysis to galaxies closer or farther from these critical points results in the assembly bias signal being attenuated in different degrees. More specifically, the main results of our work can be summarised as follows:

\begin{itemize}
\item At $z=0$, central galaxies close to nodes, filaments, and saddles are more strongly biased than galaxies far from these critical points/structures. As expected, the opposite trend is measured for voids. In general, the trends are qualitatively comparable in shape to that of assembly bias (the secondary dependence of central galaxy clustering on formation time), as they become progressively weaker towards the high-mass end.  

\item The maximum low-mass secondary bias signal is measured when splitting the galaxy population by distance to the saddle points (particularly the 1D saddles, $d_{\rm sadd1}$). The distances to nodes ($d_{\rm node}$), filaments ($d_{\rm skel}$), and voids ($d_{\rm min}$) display smaller secondary bias signals in decreasing order. For $d_{\rm sadd1}$, the difference in relative bias between same-mass subsets can exceed a value of 2 (where the relative bias is computed with respect to the entire halo mass bin). This value is significantly higher than what is measured for assembly bias ($\lesssim 0.5$). 

\item At the high-mass end, $d_{\rm sadd1}$ and $d_{\rm sadd2}$ display significant secondary bias signals, up to $\log_{10}( M_{\rm host}[h^{-1}{\rm M_\odot}]) \simeq 14$. The $d_{\rm min}$ parameter also produces secondary bias at $\log_{10}( M_{\rm host}[h^{-1}{\rm M_\odot}]) \simeq 13-13.5$. No secondary bias is found for $d_{\rm node}$ and $d_{\rm skel}$ at $\log_{10}( M_{\rm host}[h^{-1}{\rm M_\odot}]) \gtrsim 12.5-13$.

\item When analysed in pairs, selecting central galaxies by $d_{\rm node}$ and $d_{\rm skel}$ simultaneously does not alter the signal significantly, from a qualitative standpoint. However, when using $d_{\rm min}$ in combination with any other distance indicator a clear trend emerges. Central galaxies close to nodes and far from voids are more clustered than those that are just close to nodes (the inverse trend is obtained for galaxies far from nodes). Finally, simultaneously imposing the close/far-distance condition on nodes and saddles produces the largest signal measured. Here, it is also apparent that the distance to saddles dominates. 

\item Several combinations of distance parameters produce significant inversions of the signals at $\log_{10}( M_{\rm host}[h^{-1}{\rm M_\odot}]) \simeq 12.5$. The strongest crossover is measured when comparing galaxies far from nodes and close to filaments, and those sitting close to nodes and far from filaments. Below the aforementioned characteristic mass, the former are more strongly clustered, but the signal inverts above. 

\item The largest attenuation in the low-mass assembly bias signal is measured when restricting the analysis to central galaxies close to saddles and, to a lesser extent, nodes; about 34 and 41$\%$ of the signal remains at $\log_{10}( M_{\rm host}[h^{-1}{\rm M_\odot}]) \simeq 11.5$, respectively. On the contrary, the assembly bias signal remains fairly unperturbed close to voids: 83$\%$ of the signal is recovered at that halo mass. 

\item  When conditioning the clustering analysis to particular environments in the cosmic web, saddles emerge again as especial locations: not only close to them do we measure the strongest attenuation of the assembly bias signal, as mentioned above, but far from them the signal is amplified significantly (with respect to the global measurement). 

\item We have measured the redshift evolution of the secondary bias trends produced by the cosmic-web indicators for the progenitors of $z=0$ galaxies. Interesting, we find no significant evolution in any of the signals from $z=1$, as opposed to what is measured for formation time. 

\item The Pearson correlation coefficient reveals a positive, albeit modest, correlation between $d_{\rm node}$, $d_{\rm skel}$ and $d_{\rm sadd1}$ and $t_{\rm form}$ at the low-mass end, where the secondary bias signals are larger. No correlation with formation time is found for $d_{\rm min}$. 

\end{itemize}

This work presents an array of environment-related secondary clustering results that can potentially be connected with previous measurements based on internal halo properties. From a qualitative standpoint, we have shown that $d_{\rm node}$ and $d_{\rm skel}$ produce signals that resemble that of assembly bias, vanishing completely at the high-mass end. High-mass haloes are known to display strong secondary bias signals on concentration and spin (e.g., \citealt{SatoPolito2019}). It has also been discussed that other definitions of halo age could produced assembly bias signals at this range (e.g., time of last major merger, \citealt{Li2008}). Our results might indicate a certain connection between these halo properties and saddle points, which are the distance estimators that produce significant secondary bias even for the most massive haloes. Following the same argument, it is interesting to explore the relation between the density parameters that display inversions of the signal (i.e., distance to nodes) and internal halo properties such as concentration (see, e.g., \citealt{Wechsler2006, SatoPolito2019}).

As mentioned above, saddle points in the smoothed density field emerge as the best ``secondary clustering" discriminators at fixed halo mass, among the critical points provided by DisPerSE. These points correspond to local minima that are embedded inside filaments where tidal forces are expected to be significant, thus reinforcing the notion that the tidal field is an important driver of secondary bias (see, e.g., \citealt{Hahn2009,Paranjape2015,Borzyszkowski2017,Musso2018, Paranjape2018, Ramakrishnan2019}). In particular, we could speculate that the fact that the assembly bias signal is severely attenuated close to saddles and is actually enhanced far from them (Figs. \ref{fig:AB3} and \ref{fig:percentage}) could be related to the abundance of stalled and accreting haloes. It is plausible that by restricting the measurement to objects close to saddles, we are effectively targeting a population of (mostly) stalled, non-accreting haloes. Conversely, when the measurement is opened to any haloes far from saddles, we could be mapping a broader halo population. Follow-up work will be devoted to address this point in detail, and, particularly, to establish connections with previous analyses that have highlighted the relevance of saddle points in the cosmic web (e.g., \citealt{Musso2018, Codis2015}). 

Our work can be considered complementary to the analysis of the connection between the anisotropy parameter, $\alpha$, and secondary bias  (\citealt{Paranjape2018} and thereafter). In those works, the secondary bias signals are claimed to emerge from correlations between halo properties and $\alpha$, and between $\alpha$ and halo bias, at fixed halo mass. As follow-up work, evaluating in detail the connection between $\alpha$ and the DisPerSE parameters in the context of assembly bias could provide additional clues to the physical origins of the effect. 

Importantly, the DisPerSE method has already been applied to galaxy surveys \citep[e.g.,][]{Luber2019} and so have other algorithms for identifying structures in the cosmic web. These datasets include catalogues of filaments \citep[e.g.,][]{Tempel2014, Martinez2016, Pereyra2020}, voids \citep[e.g.,][]{Ruiz2015, Paz2023} and groups and clusters of galaxies \citep[e.g.,][]{Tempel2017, Rodriguez2020}. Since some of the secondary bias signals found in this work are significant up to halo masses of $\log_{10}( M_{\rm host}[h^{-1}{\rm M_\odot}]) \simeq 12.5$ and above, it is interesting to explore, as follow-up work, whether similar clustering results can be found in observations.

\section*{Acknowledgments}

Most of this work was carried out during our visit, as research associates, to the Abdus Salam International Centre for Theoretical Physics (ICTP), in the summer of 2023. ADMD and FR thank the ICTP for their hospitality and financial support through the Senior Associates Programme 2022-2027 and Junior Associates Programme 2023-2028, respectively. We also thank Ravi K. Sheth for inspiring discussions during our visit. 

ADMD thanks Fondecyt for financial support through the Fondecyt Regular 2021 grant 1210612. FR thanks the support by Agencia Nacional de Promoci\'on Cient\'ifica y Tecno\'ologica, the Consejo Nacional de Investigaciones Cient\'{\i}ficas y T\'ecnicas (CONICET, Argentina) and the Secretar\'{\i}a de Ciencia y Tecnolog\'{\i}a de la Universidad Nacional de C\'ordoba (SeCyT-UNC, Argentina).

\section*{Data availability}
The simulation data underlying this article are publicly
available at the TNG website. The data results arising from this work will be shared on reasonable request to the corresponding authors.

\bibliographystyle{mnras}
\bibliography{biblio}

\label{lastpage}

\end{document}